\documentclass{article}

\usepackage{PRIMEarxiv}

\usepackage[utf8]{inputenc} % allow utf-8 input
\usepackage[T1]{fontenc}    % use 8-bit T1 fonts
\usepackage{hyperref}       % hyperlinks
\usepackage{url}            % simple URL typesetting
\usepackage{booktabs}       % professional-quality tables
\usepackage{amsfonts}       % blackboard math symbols
\usepackage{nicefrac}       % compact symbols for 1/2, etc.
\usepackage{microtype}      % microtypography
\usepackage{lipsum}
\usepackage{fancyhdr}       % header
\usepackage{graphicx}       % graphics
\graphicspath{{media/}}     % organize your images and other figures under med\cite{}ia/ folder

\usepackage{tabularx}       % resize table
\usepackage{multirow}
\usepackage{xcolor}

\title{Anatomic Feature Fusion Model for Diagnosing Calcified Pulmonary Nodules on Chest X-ray
}

\author{
  Hyeonjin Choi \\
  Affiliation \\
  Ajou University \\
  Suwon City\\
  \texttt{hjin9122@gmail.com} %%{\{Author1, Author2\}email@email} \\
   \And
  Yang-gon Kim  \\
  Affiliation \\
  Ajou University \\
  Suwon City\\
  \texttt{didrhs96@gmail.com} \\
   \And
  Dong-yeon Yoo  \\
  Affiliation \\
  Ajou University \\
  Suwon City\\
  \texttt{dongs0125@ajou.ac.kr} \\
  \AND
  Ju-sung Sun \\
  Affiliation \\
  Ajou University Hospital \\
  Suwon City\\
  \texttt{sunnahn@ajou.ac.kr} \\
  \And
  Jung-won Lee \\
  Affiliation \\
  Ajou University \\
  Suwon City\\
  \texttt{jungwony@ajou.ac.kr} \\
}

\begin{document}
\maketitle

\begin{abstract}
    Accurate and timely identification of pulmonary nodules on chest X-rays can differentiate between life-saving early treatment and avoidable invasive procedures. Calcification is a definitive indicator of benign nodules and is the primary foundation for diagnosis. In actual practice, diagnosing pulmonary nodule calcification on chest X-rays predominantly depends on the physician’s visual assessment, resulting in significant diversity in interpretation. Furthermore, overlapping anatomical elements, such as ribs and spine, complicate the precise identification of calcification patterns. This study presents a calcification classification model that attains strong diagnostic performance by utilizing fused features derived from raw images and their structure-suppressed variants to reduce structural interference. We used 2,517 lesion-free images and 656 nodule images (151 calcified nodules and 550 non-calcified nodules), all obtained from Ajou University Hospital. The suggested model attained an accuracy of 86.52\% and an AUC of 0.8889 in calcification diagnosis, surpassing the model trained on raw images by 3.54\% and 0.0385, respectively. 
\end{abstract}

% keywords can be removed

\section{Introduction}
    The diagnosis of Pulmonary Nodule(PN) should ideally reduce the number of needless procedures (e.g., biopsies) for benign nodules while facilitating therapy for malignant nodules \cite{patel2013practical, wahidi2007evidence, comstock1956outcome}. Calcification of PNs is a definitive indicator of benignity and serves as the primary criterion for diagnosing benignity versus malignancy \cite{zhou2021calcification, winer2006solitary, erasmus2000solitary, adebonojo1975evaluation, good1958solitary}. If a PN has a benign calcification pattern (e.g., diffuse, central, laminar, popcorn), it is classified as benign. Conversely, if a PN exhibits no calcification or, infrequently, a calcification pattern of ambiguous benign/malignant nature (e.g., Eccentric, Small Flecks), it is presumed to be malignant. PNs are frequently identified on chest X-ray(CXR) and diagnosed with Computed Tomography(CT) scans for calcification assessment \cite{panunzio2020lung}. CT imaging is useful for detecting patterns within nodules precisely because it provides a three-dimensional chest image \cite{furtado2005whole, ea1986ct}. Although CT is valuable, it is costly and poses a significant risk of elevated radiation exposure \cite{fatihoglu2016x, smith2009radiation}. Thus, the importance of accurate diagnosis during the CXR phase, which is economical and employs minimal radiation, has been emphasized. 
    
\begin{figure}
      \centering
      \includegraphics[width=\textwidth]{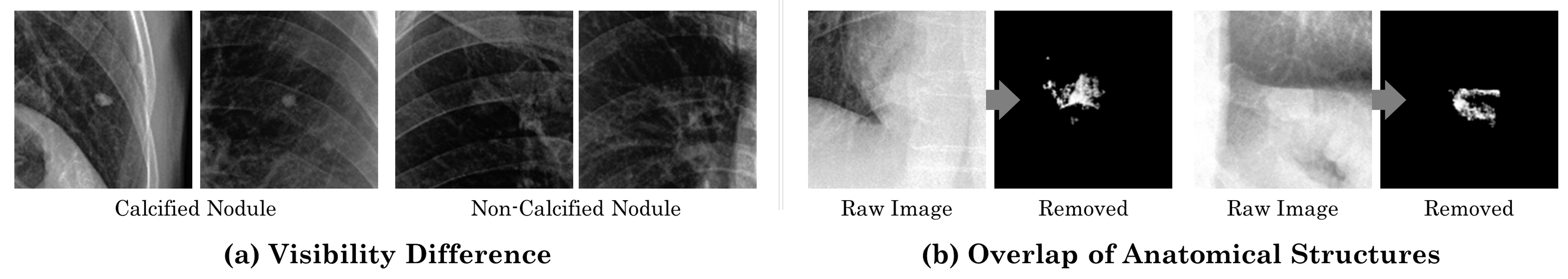}
      \caption{\textbf{Examples illustrating the challenges in nodule diagnosis.} (a) demonstrates the visible disparities between calcified and non-calcified nodules, with calcified nodules appearing more pronounced. (b) illustrates the enhancement in nodule visibility upon the removal of overlapping anatomical features.}
      \label{fig:fig1}
    \end{figure}

    In practice, however, accurately identifying PNs on CXR is a significant challenge. One reason for this challenge is that physicians' identification of nodule calcification on CXRs depends on visual assessment. Hence, the physician's experience or exhaustion, along with the characteristics of the nodules, may result in variations in diagnostic outcomes \cite{blanchon2007baseline, mahoney1990ct}. For instance, the nodule's size affects how well calcification is diagnosed. Small-sized nodules (under 10 mm in diameter) identified on a CXR are more likely to exhibit calcification \cite{you2019visual}. Non-calcified nodules are challenging to locate when small because of their low brightness; however, they are more likely to be calcified if they have a discernible brightness. Fig.~\ref{fig:fig1}(a) illustrates the visibility of a nodule of identical size contingent upon its calcification status. Another reason is that the anatomical features, such as ribs and vertebrae, superimpose on the nodule, complicating its exact identification \cite{berger2001solitary}. In Fig.~\ref{fig:fig1}(b), the raw image makes it challenging to ascertain the existence of a nodule. Yet the anatomy-removed image verifies its presence. Therefore, to enhance diagnosis accuracy for PNs in CXRs, a diagnostic assistance tool that is unaffected by these variables and can deliver consistent diagnosis performance is required.
    
    \begin{figure}
      \centering
      \includegraphics[width=\textwidth]{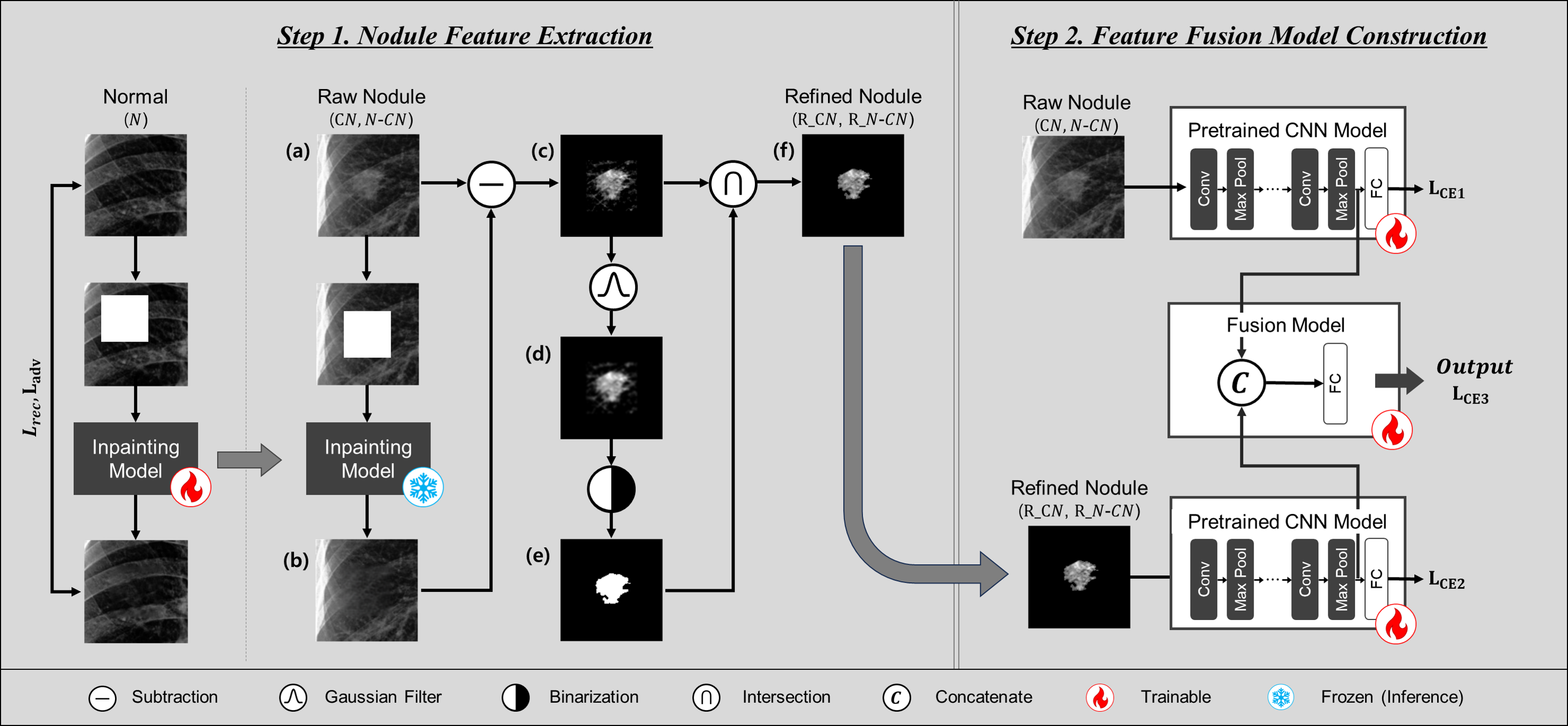}
      \caption{\textbf{Process of Constructing Anatomic Feature Fusion Model.} The left depicts the initial phase, whereby an inpainting model trained on normal set and subsequent operations are employed to derive refined set from nodule set. The right illustrates the second stage, which integrates nodule and refined set to formulate the feature fusion model.}
      \label{fig:fig2}
    \end{figure}
    
    AI-based diagnostic assistance tools have effectively diagnosed PN \cite{megat2025systematic, khan2024ai, niehoff2023evaluation, yoo2021ai}. Current methodologies involve eliminating anatomical components or enhancing lesion characteristics \cite{chu2024learning, gaggion2024chexmask, kamal2022anatomy, kondo2021chest} and image preprocessing or feature extraction from the target lesion for fusion learning \cite{vats2024iterative, agarwal2024multifusionnet, sumathi2024efficient, paul2018predicting}. However, from the perspective of clinical application, most current studies define malignant PNs as lung cancer and only seek to determine whether they are lung cancer or not. This is inefficient because it fails to consider other lesions that malignant PNs may have, such as lung abscesses, tuberculosis, etc. In this work, we suggest a feature fusion-based calcification diagnosis model that can help with diagnosis broadly and is not restricted to a particular lesion.

\section{Methods}
\label{method}
    Constructing an anatomical feature fusion model for diagnosing nodule calcification 
    involves two stages: nodule feature extraction and fusion learning (Fig.~\ref{fig:fig2}).
    
    \subsection{Nodule Feature Extraction}
        The nodule feature extraction process seeks to identify nodule regions within a nodule dataset (Calcified Nodule: CN, Non-Calcified Nodule: N-CN) and to extract nodule features exclusively from those regions. But because nodule areas are so similar in brightness to the surrounding structures, it is hard to tell them apart from the naked eye. This means the acquisition of nodule mask data is challenging, rendering the application of segmentation methods unfeasible. Thus, this study employs inpainting to eliminate the nodule regions from the raw image and extract the nodule areas by calculating the difference between the modified and original images. Fig.~\ref{fig:fig2} illustrates the comprehensive procedure for nodule feature extraction in the left region. 
    
        \begin{itemize}
            \item[1.] \textbf{Nodule Removal and Anatomic Structure Restoring with Inpainting. (Fig.~\ref{fig:fig2}(a)\&(b))} Inpainting is a method for restoring damaged or absent sections of an image \cite{elharrouss2020image}. Initially created to restore corrupt images, it may also eliminate uncommon features inside an image. An inpainting model trained exclusively on normal(lesion-free) data (N) can accurately reconstruct anatomical structures. However, because this dataset contains no information on nodules, the model cannot discern the characteristics of the nodules within the image and, consequently, cannot reconstruct their appearance, i.e., exclude them from the nodule image. Leveraging this property, this paper delineates a procedure to generate a nodule-removed image (Fig.~\ref{fig:fig2}(b)) by masking the central region of a nodule dataset (CN/N-CN) and entering it into an inpainting model trained on normal data. 
        
            \item[2.] \textbf{Nodule Extraction by Subtraction. (Fig.~\ref{fig:fig2}(c))} By executing a subtraction operation between the raw nodule image (Fig.~\ref{fig:fig2}(a), CN/N-CN) and the inpainting result (Fig.~\ref{fig:fig2}(b)), we get an image containing just the feature of the nodule (Fig.~\ref{fig:fig2}(c)), in which the features of anatomical structures have been eliminated. 
        
            \item[3.] \textbf{Nodule Region Determination by Denoising. (Fig.~\ref{fig:fig2}(d)\&(e))} The previously extracted nodule feature (Fig.~\ref{fig:fig2}(c)) is contaminated by noise from the CXR image and the inpainting procedure, which impedes diagnosis. To eliminate this, we isolate the feature relating to the nodule region (Fig.~\ref{fig:fig2}(e)). A Gaussian filter of size 5x5 is initially applied to extracted nodule features to decrease the noise. Fig.~\ref{fig:fig2}(d) depicts an image that has been denoised using a Gaussian filter. Subsequently, we implement binarization using the Otsu algorithm to delineate the nodule's region by removing noise from the previous process and improving the unclear contours \cite{goh2018performance}. 
        
            \item[4.] \textbf{Nodule Filtering by Intersection (Fig.~\ref{fig:fig2}(f))} The intersection operation is performed on the nodule information obtained via subtraction (Fig.~\ref{fig:fig2}(c)) and the area information of nodules identified post-denoising (Fig.~\ref{fig:fig2}(e)). This procedure yields denoised nodules (Fig.~\ref{fig:fig2}(f)), which ultimately constitutes the Refined Nodule (Refined Calcified Nodule: R\_CN/ Refined Non-Calcified Nodule: R\_N-CN) image to be acquired in the Nodule Feature Extraction phase. 
        
        \end{itemize}
            
    As a result, the anatomical structures and noise around the nodule were removed, al
    lowing only clear nodule information to be obtained 
        
    \subsection{Feature Fusion Model Construction}
        \textbf{Feature Fusion Deep Learning.} Fusion Learning is conducted utilizing pairs of nodule datasets to construct a model that reflects the characteristics of PNs as indicated by both the refined nodule data(R\_CN/R\_N-CN) and the raw nodule image(CN/N-CN). The architecture of the feature fusion-based calcification diagnosis model is illustrated on the right side of Fig.~\ref{fig:fig2} The process involves extracting deep features from raw nodule data and refined nodule data, followed by estimating the presence of calcification based on the fused deep features. In this context, deep features refer to the outputs of the final pooling layer of each CNN model. The two CNN models and the final classifier function independently, each with distinct loss(L\_(CE1-3)). The CNN models are trained to reduce calcification prediction loss the L\_CE1 and L\_CE2 from individual input, while the final classifier is trained to minimize the L\_CE3 from the integrated features. 
    
\section{Experiments}
\label{Experiments}

    \subsection{Dataset and Processing}
        \textbf{Dataset.} This study utilized 3,173 chest X-ray (CXR) images (2,517 normal, 656 with nodules) and corresponding radiographic records from  Ajou University Hospital, Suwon, Republic of Korea. The radiographic records included the number of nodules depicted, locations and size, the presence of calcification, and the subjective clarity of the nodules as assessed by the radiologist. This study used solely CXR images and nodule location, size, and calcification status.
        
        \textbf{Data Preprocessing.} Initially, we extract the area surrounding the nodule and specific regions within the lung from the image, subsequently creating a nodule dataset and a normal dataset, respectively. We will refer to the raw CXR images as “Full View (FV)” to eliminate ambiguity, as we exclusively use the cropped dataset for model development. The nodule dataset (CN, N-CN) is derived by cropping the center coordinates of the nodule with uniform edges applied above, downward, and laterally. The normal dataset (N) is constructed by clipping a random lung location from a nodule-free FV CXR image, using center coordinates that match the size of the nodule dataset. Cropping Size (CS) was selected to encompass FV CXR's largest nodules, while the Resizing Ratio (R) was set for image processing efficiency. The Boxing Lung Region ($B$)) for extracting the normal dataset was established, considering that the left and right sides encompass external body areas, while the bottom includes a significant portion of the organs below. The Number of Randomly Selected Coordinates ($C$) for each image was assigned a random value. To generate significant data, the coordinates are designed to change while allowing cropped image overlaps. Table.~\ref{tab:table1} displays the values for each setting. The final datasets comprise 550 CN, 151 N-CN, and 45,288 N. 
        
        \textbf{Data Augmentation.} We augment the calcification dataset (CN, R\_CN) because calcified nodule images account for only 27.5\% of the overall nodule images, resulting in a significant imbalance. The same augmentation strategy was applied to the paired raw and refined nodule data. It was enhanced four times over the original by translating it up, down, left, and right and then applying Horizontal Flip, Rotation, Aspect Ratio Adjustment, and Resolution Adjustment. The final feature fusion training dataset is shown in Table.~\ref{tab:table3} after randomly applying augmentation parameters (Table.~\ref{tab:table2}). 

    \newcolumntype{Y}{>{\centering\arraybackslash}X} % auto center alignment
    \begin{table}
     \caption{Experimental Setting’s Parameters}
      \centering
      \begin{tabularx}{\textwidth}{cYc}
        \toprule
        \qquad Parameters \qquad \qquad     & Detail     & Value \\
        \midrule
        $CS$    & Cropping Size (width x height)  & 512 x 512 pixels    \\
        $R$     & Resizing Ratio & 1/4      \\
        $C$     & Number of Random Selected Coordinates & 20  \\
        $B$     & Boxing Lung Region (x, y, width, height) & \qquad 192, 192, 1664, 1344 pixels \qquad \\
        \bottomrule
      \end{tabularx}
      \label{tab:table1}
    \end{table}
    
    \begin{table}
      \caption{Range of Random Data Augmentation Parameters}
      \centering
      \begin{tabularx}{\textwidth}{YY}
        \toprule
        Parameter & Range \\
        \midrule
        Translation & $x\pm32, y\pm32$ \\
        Horizontal Flip & $p=0.5$ \\
        Rotation & $[-18^\circ, +18^\circ]$ \\
        Aspect Ratio Adjustment & $[0.75, 1.25]$ \\
        Resolution Adjustment & $[0.75, 1.25]$ \\
        \bottomrule
      \end{tabularx}
      \label{tab:table2}
    \end{table}
    
    \begin{table}
     \caption{Dataset Composition for Fusion Learning}
      \centering
        \begin{tabularx}{\textwidth}{YYYYY}
        \toprule
        \qquad & \multicolumn{2}{c}{Calcified} & \multicolumn{2}{c}{Non-Calcified} \\
              & CN           & R\_CN          & N-CN           & R\_N-CN          \\ \midrule
        Train & 480          & 480            & 440            & 440              \\
        Valid & 31           & 31             & 110            & 110              \\ \midrule
        Total & \multicolumn{2}{c}{1,022}     & \multicolumn{2}{c}{1,100}         \\ \toprule
        \end{tabularx}
        \label{tab:table3}
    \end{table}
    
    \subsection{Experimental Setup.}
        \textbf{Inpainting Model Construction.} We employed an inpainting model based on the architecture of Context Encoder \cite{chen2024context}. It comprises six 4x4 convolutional layers and five 4x4 up-convolutional layers. The adversarial discriminative model for adversarial loss learning shall consist of five 4x4 convolutional layers. The Loss function ($L_{inpaint}$) is defined as a weighted sum of the reconstruction loss ($L_{rec}$) and the adversarial loss ($L_{adv}$) to address both local and global discrepancies between the original and restored images. The arbitrary parameters($L_{rec}, L_{adv}$) are employed to establish the weighting between the losses, with their sum equaling (1). 
    
        \begin{equation}
            L_{inpaint} = \lambda_{rec}L_{rec}+\lambda_{adv}L_{adv}
        \end{equation}
    
        \textbf{Inpainting Model Training.} A normal dataset (N) is employed to train the model. The masking region's position is randomized, and its size is established at 20\% of the overall image, which was experimentally determined to yield satisfactory restoration performance. The model parameters are configured with the Adam Optimizer, a batch size 16, a learning rate of 0.0002, and trained for 500,000 epochs. The parameters of the loss function were configured to $L_{rec}=0.999$ and $L_{adv}=0.001$. The performance of the inpainting model during training can be assessed using the Mean Square Error (MSE) and Peak Signal-to-Noise Ratio (PSNR) by comparing the restoration results to the original data on a dataset not utilized for training. The two measures have an inverse relationship; hence, as the MSE approaches zero and the PSNR value increases, the restoration becomes increasingly like the original image. The inpainting model developed in this study had superior restoration efficacy, evidenced by a low MSE of 0.0036 and a high PSNR of 20.60, and was employed for nodule elimination.

    \subsection{Implementation Details.} 
        The two CNN models in the feature fusion learning framework employed ResNet-50 \cite{he2016deep}. pre-trained on ImageNet \cite{deng2009imagenet}. The model parameters were configured with the Adam Optimizer, Cross Entropy Loss Function, a batch size 32, a learning rate of 0.00001, and 100 training epochs

\section{Results}
\label{Results}

    \begin{figure}
      \centering
      \includegraphics[width=0.95\linewidth]{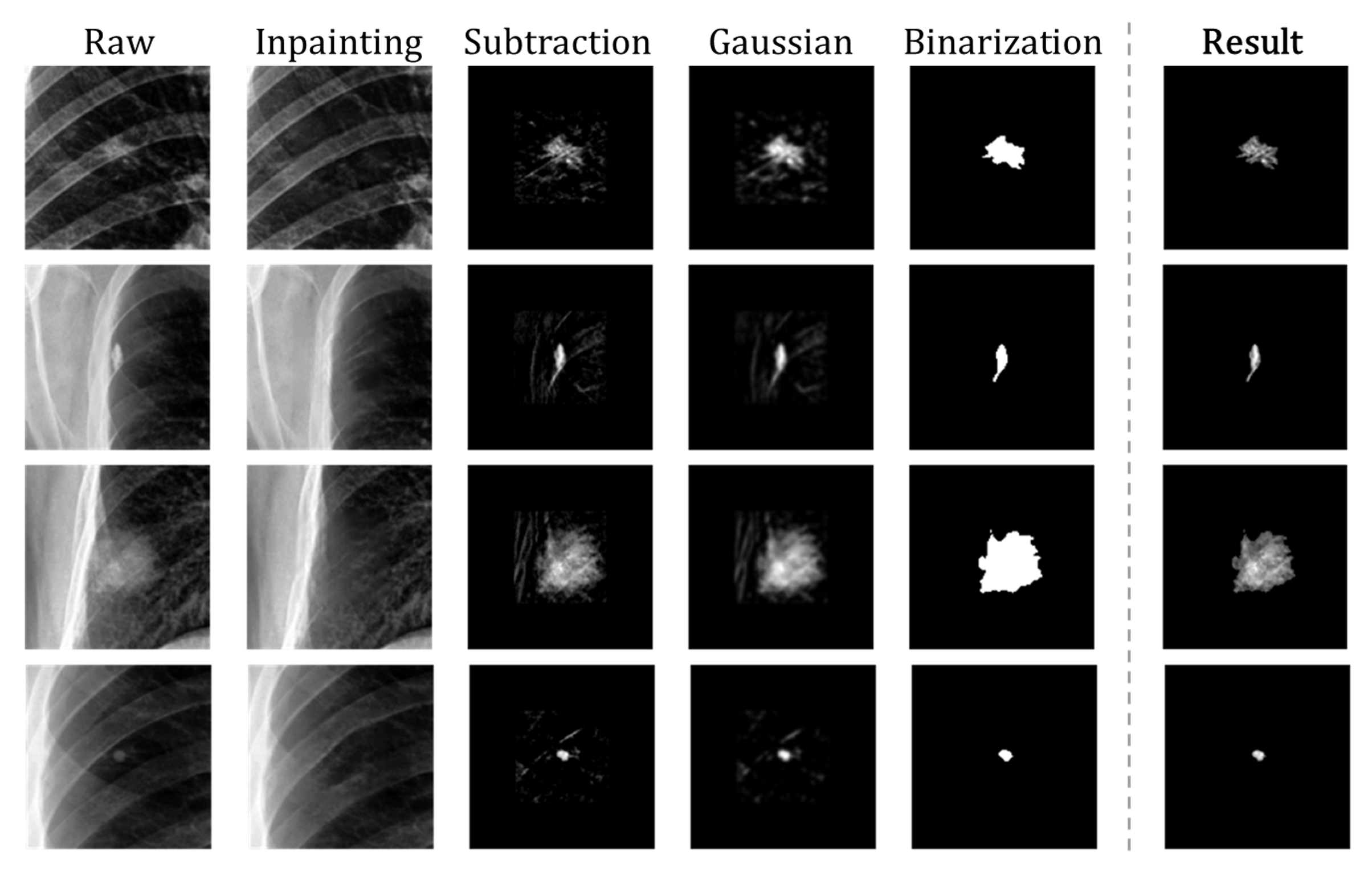}
      \caption{Example of the Result of the Pulmonary Nodule Feature Extraction Process}
      \label{fig:fig3}
    \end{figure}
    
    \begin{figure}
      \centering
      \includegraphics[width=0.75\linewidth]{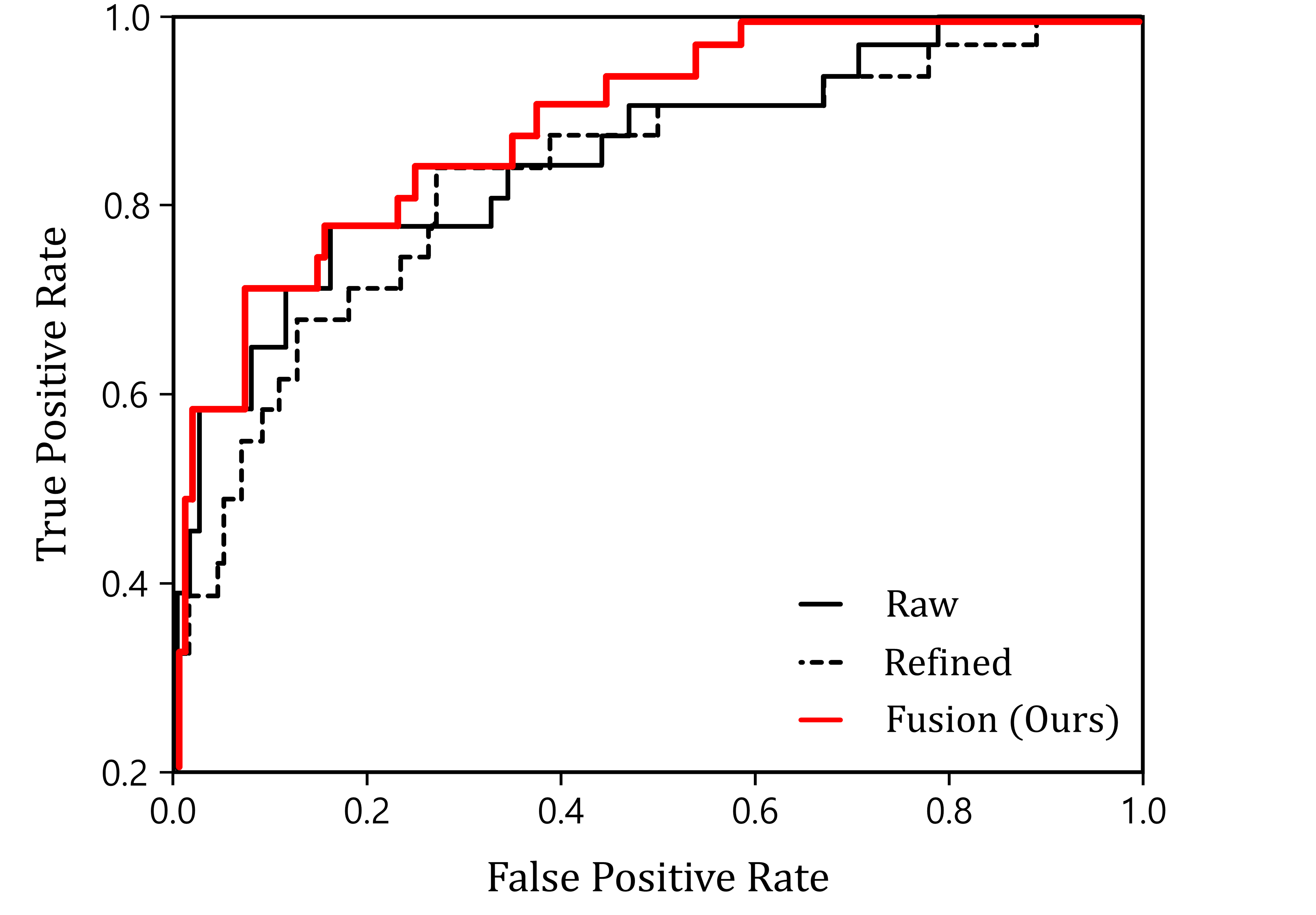}
      \caption{ROC Curves of Single Models(Raw, Refined) and Our Proposed Fusion Model.}
      \label{fig:fig4}
    \end{figure}
    
    \begin{table}
      \caption{Performance of the Feature Fusion Model }
      \centering
      \begin{tabularx}{\textwidth}{YYYY}
      \toprule
        \multirow{2}{*}{Model}  & \multirow{2}{*}{Input Type}  & \multicolumn{2}{c}{Validation} \\
                                &                              & Accuracy       & AUC           \\ \midrule
        Single                  & Raw                          & 0.8298         & 0.8504        \\
        \multirow{2}{*}{Single} & \multirow{2}{*}{Refined}     & 0.8085         & 0.8305  \rule{0pt}{3ex}      \\
                                &                              & (-2.13\%)      & (-1.99\%)     \\
        \multirow{2}{*}{\textbf{Fusion}} & \multirow{2}{*}{\textbf{Raw+Refined}} & \textbf{0.8652}         & \textbf{0.8889}  \rule{0pt}{3ex}      \\
                                &                              & \textcolor{red}{(+3.54\%)}      & \textcolor{red}{(+3.85\%)}     \\ \toprule 
      \end{tabularx}
      \label{tab:table4}
    \end{table}
    
    Fig.~\ref{fig:fig3} illustrates the sequential outcomes of the nodule feature extraction methodology introduced in this research. The inpainting results indicate that the nodules in the image have been eliminated, allowing for the precise extraction of nodule-specific features without any noise. The final learning outcome was assessed using accuracy, the Area Under Curve (AUC) score, and the Receiver Operating Characteristic (ROC) curve as metrics (Table.~\ref{tab:table4}, Fig.~\ref{fig:fig4}). The efficacy of the single model trained on the refined dataset (R\_CN, R\_N-CN) was about 2\% lower than that of the single model trained on the raw dataset (CN, N-CN). This can be related to losing data during the extraction of refined nodule data. The feature fusion model, which was trained on both the raw and the refined nodule dataset, exhibited an enhancement of 3.54\% accuracy and 0.0385 in AUC relative to the raw images alone. Furthermore, the ROC Curve indicates that the feature fusion model surpassed the single model in terms of recall and specificity. The results suggest that the distinct characteristics of the raw and refined images are complementary and constitute the primary basis for diagnosing calcification. 

\section{Conclusion}
    This paper presents a deep learning model utilizing feature fusion to assist physicians in determining the calcification of PNs in CXRs, aiming to enhance diagnostic accuracy for such lesions. The suggested model employs an enhanced nodule extraction method and a feature fusion learning approach. The refined nodule extraction method isolates nodule-specific features concealed by anatomical structures, strengthening the model's sensitivity to detect benign nodule indicators. The feature fusion learning method supplies the model with both the raw and refined nodule images to prevent losing data that could occur during image refinement. 
    
    The proposed feature fusion learning method demonstrated superior results to the single-image learning model. Also, it showed better efficacy in diagnosing calcification, achieving an AUC of 0.8889, surpassing the diagnostic AUC of 0.835 observed in the clinic. The proposed model serves as a diagnostic tool, providing a secondary opinion for physicians in the clinic by effectively identifying and suggesting calcification with high accuracy. Furthermore, in contrast to current studies that solely identify the presence of lung cancer, the proposed model can indicate the presence of calcification, which serves as a primary diagnostic criterion, enhancing diagnostic support more comprehensively. 
    
    This study was subject to limitations in conducting a comprehensive generalization test due to constraints in data availability \cite{kim2019short}.  Publicly available datasets annotated with calcification status remain limited, which poses challenges for further validation. Future work will address these limitations by expanding data augmentation techniques and acquiring additional external datasets to enhance generalizability and reproducibility. 

% \ref{sec:headings}.

\section*{Acknowledgments}
This work was supported by the IITP(Institute of Information \& Coummunications Technology Planning \& Evaluation)-ITRC(Information Technology Research Center) grant funded by the Korea government(Ministry of Science and ICT)(IITP-2025-RS-2020-II201461)

%Bibliography
\bibliographystyle{unsrt}  
\bibliography{references}

\end{document}